\documentclass[twocolumn,showpacs,preprintnumbers,amsmath,amssymb, superscriptaddress, prl]{revtex4}

\usepackage{graphicx}
\usepackage{dcolumn}
\usepackage{bm}
\newcommand{\FeTe} {{\mbox{FeTe${}_2$O${}_5$Br }}}
\newcommand{\half}{{\ensuremath{\frac{1}{2}}}}
\newcommand{\thalf}{{\ensuremath{\frac{3}{2}}}}
\usepackage{color}
\definecolor{red}{rgb}{1,0,0}

\definecolor{blue}{rgb}{0,0,1}

\definecolor{green}{rgb}{0,1,0}

\begin{document}
\preprint{APS}

\title{Spin amplitude modulation driven magnetoelectic coupling in the new multiferroic \FeTe}
\author{M. Pregelj}
\affiliation{Institute "Jozef Stefan", Jamova 39, 1000 Ljubljana, Slovenia}
\author{O. Zaharko }
\email{oksana.zaharko@psi.ch}
\affiliation{Laboratory for Neutron Scattering, ETHZ \& PSI, CH-5232 Villigen, Switzerland}
\author{A. Zorko}
\affiliation{Institute "Jozef Stefan", Jamova 39, 1000 Ljubljana, Slovenia}
\author{Z. Kutnjak}
\affiliation{Institute "Jozef Stefan", Jamova 39, 1000 Ljubljana, Slovenia}
\author{P. Jegli\v{c}}
\affiliation{Institute "Jozef Stefan", Jamova 39, 1000 Ljubljana, Slovenia}
\author{P. J. Brown}
\affiliation{Institut Laue-Langevin, 156X, 38042 Grenoble Cedex, France}
\author{M. Jagodi\v{c}}
\affiliation{Institute of Mathematics, Physics and Mechanics, Jadranska 19, 1000 Ljubljana, Slovenia}
\author{Z. Jagli\v{c}i\' {c}}
\affiliation{Institute of Mathematics, Physics and Mechanics, Jadranska 19, 1000 Ljubljana, Slovenia}
\affiliation{Faculty of Civil and Geodetic Engineering, University of Ljubljana, Jamova 2, 1000 Ljubljana, Slovenia}
\author{H. Berger}
\affiliation{Institute of Physics of Complex Matter, EPFL, 1015 Lausanne, Switzerland}
\author{D. Ar\v{c}on}
\email{denis.arcon@ijs.si}
\affiliation{Institute "Jozef Stefan", Jamova 39, 1000 Ljubljana, Slovenia}
\affiliation{Faculty of Mathematics and Physics, University of Ljubljana, Jadranska 19, 1000 Ljubljana, Slovenia}
\date{\today}

\begin{abstract}
Magnetic and ferroelectric properties of layered geometrically frustrated cluster compound \FeTe  were investigated with single-crystal neutron diffraction and dielectric measurements. Incommensurate transverse amplitude modulated magnetic order with  the wave vector $\bf{q}$=($\half$, 0.463, 0) develops below $T_N=10.6(2)\, {\rm K}$. Simultaneously, a ferroelectric order  due to exchange striction involving polarizable Te$^{4+}$ lone-pair electrons develops perpendicular to ${\bf q}$ and to Fe$^{3+}$ magnetic moments. { The observed magnetoelectric coupling is proposed to originate from the temperature dependent phase difference between neighboring amplitude modulation waves. }
\end{abstract}

\pacs{75.25.+z, 75.80.+q}
\maketitle

Switching ferroelectric polarization by magnetic field \cite{Kimura} or, conversely, controling magnetic order with the electric field \cite{Lottermoser} in magnetoelectric materials has been for a long time hampered by a very small magnitude of the magnetoelectric coupling. Recently, strong magnetoelectric coupling has been discovered in several multiferroic oxides ($R$MnO$_3$, $R$Mn$_2$O$_5$, Ni$_3$V$_2$O$_8$, $\ldots$ , $R$ = rare earth) where ferroelectricity exists only in a magnetically ordered state \cite{Nmat07, Fiebig05, ScottNature, Kimura, Lottermoser}. In these systems, spiral magnetic order, such as cycloidal or transverse conical structures \cite{Kasuga}, breaks the inversion symmetry and removes  strict symmetry restrictions for the existence of the magnetoelectric coupling. Since spiral order often results from magnetic frustration, the current focus is on materials with geometrically frustrated lattices.
However, it remains to be seen whether the strong magnetoelectric effect is restricted to spiral magnetic structures or it can  be found also in other spatially modulated magnetic arrangements.

In recent years several geometrically frustrated spin-cluster oxyhalide compounds $M$-Te-O-$X$ ($M$ = Cu, Ni, Fe; $X$ = Cl, Br, I) have been synthesized. Because of their reduced magnetic dimensionality and frustrated lattices they frequently exhibit a complex magnetic order, having low magnetic symmetry. Moreover, these systems contain Te$^{4+}$ ions with lone-pair electrons ($5s^25p^0$), which are highly polarizable \cite{LonePair}. Thus the $M$-Te-O-$X$ family represents a new class of materials, where magnetic and polar order may coexist. We have focused our investigations  on  FeTe$_2$O$_5$Br  with a crystal structure that implies magnetic frustration \cite{Becker}. This system crystallizes in a monoclinic unit cell (space group $P21/c$) and adopts a layered structure. The layers, which are stacked along the crystal $a^*$-axis, consist of triangularly arranged [Fe$_4$O$_{16}$]$^{20-}$ clusters linked by [Te$_4$O$_{10}$Br$_2$]$^{6-}$ units. Within each iron tetramer cluster there are two crystallographically non-equivalent Fe$^{3+}$ ($S$ = 5/2) ions (Fe1 and Fe2 on 4($e$) sites) coupled through competing antiferromagnetic exchange interactions. In this letter we show that below the Neel transition temperature $T_N=10.6\, {\rm K}$ the Fe$^{3+}$ magnetic moments order almost collinearly with an incommensurate amplitude modulation. A spontaneous electric polarization associated with the polarizable Te$^{4+}$ lone-pair electrons appears simultaneously with the long-range magnetic order.
{ We propose that the phase difference between coupled modulation waves is responsible for the magnetoelectric effect in FeTe$_2$O$_5$Br and possibly also in other incommensurate amplitude modulated magnetic structures.}

Single crystals of FeTe$_2$O$_5$Br were grown by standard chemical vapor phase method. Single-crystal X-ray diffraction measurements ($\lambda =$0.64~\AA) were performed at the BM01A Swiss-Norwegian Beamline of ESRF, France using closed-cycle He cryostat mounted on a six-circle kappa diffractometer KUMA. Data sets collected in the temperature range 4.5 to 35 K were refined using the  SHELXL program \cite{SHELXL}. Neutron integrated intensities were collected on a $5 \times 4 \times 1$ mm$^3$ single crystal at 5 K on the single crystal diffractometer
TriCS ($\lambda$ =2.32~\AA) at the Swiss Neutron Spallation Source, Switzerland. Spherical neutron polarimetry measurements on a $7 \times 5 \times 1.6$ mm$^3$ single crystal were carried out at 1.8 K with CRYOPAD II installed on the IN20 spectrometer ($\lambda$=2.34~\AA) at the Institute Laue-Langevin, France. The crystal was mounted with the $c$-axis perpendicular to the scattering plane. The complex dielectric constant $\epsilon^*(T, \omega)= \epsilon'(T,\omega)-i\epsilon'' (T,\omega)$ was measured as a function of temperature and frequency with the HP4282A precision LCR meter. The quasistatic polarization $P$ was determined by electrometer charge accumulation measurements \cite{eps1, eps2} in a field cooling run (a bias field of 10 kV/cm).

\begin{table}[t]
\caption{
Neutron polarization matrices $P_{ij}$ ($i$ - incoming,  $j$ - outcoming component of
polarization) for two representative reflections measured at $T=1.8\, {\rm K}$.
\label{tab1}}
\begin{ruledtabular}
\begin{tabular}{ccccccc}
\multicolumn{3}{c}{${h~~~~~~~~k~~~~~~~~l}$}&${P_{i}}$&${P_{ix}}$&${P_{iy}}$&${P_{iz}}$\\
${\half}$&-0.463&0&$x$&-0.85(2)&0.05(1)&0.04(1)\\
&&&$y$&0.03(1)&0.83(1)&-0.09(1)\\
&&&$z$&-0.00(1)&-0.10(1)&-0.77(1)\\
\\
${\thalf}$&1.537&0&$x$&-0.927(4)&0.05(1)&0.01(1)\\
&&&$y$&0.01(1)&0.823(6)&0.34(1)\\
&&&$z$&-0.04(1)&0.38(1)&-0.843(6)\\
\end{tabular}
\end{ruledtabular}
\end{table}

\begin{figure}[b]
\includegraphics[width=80mm,keepaspectratio=true]{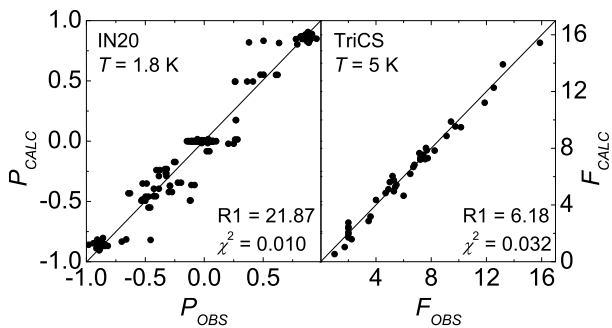}
\caption {The agreement between experimental and calculated quantities E: (left) components of neutron polarization matrices E=$P$ and (right) magnetic structure factors E=$F$. The reliability factors are defined as: R1=$\Sigma \Delta E/\Sigma E$ and $\chi^2=(\Delta E)^2/(N_{observables}-N_{parameters})$.}
\label{fig1}
\end{figure}

{In \FeTe three-dimensional long-range magnetic ordering sets in at $T_N = 10.6(2)\, {\rm K}$, where a pronounced change in the temperature dependence of $\chi$ is evident \cite{Becker}. Our neutron diffraction measurements reveal that the magnetic reflections emerge at the incommensurate positions described by the wave vector $\bf{q}$=($\half$, 0.463, 0).
Close inspection of polarization matrices obtained from neutron polarimetry measurements (Table~\ref{tab1}) indicates that the magnetic arrangement is neither a spiral, nor a cycloid or strongly canted. The absence of the $P_{yx}$ and $P_{zx}$ components and almost full polarization of the scattered beam implies that chiral magnetic scattering is negligible. The off-diagonal components $P_{yz}$ and $P_{zy}$ increase  with increasing $h$ or $k$ suggesting a small $c$-component of magnetic moment.}

\begin{table} [t]
\caption{Parameters of the magnetic structure deduced from neutron diffraction experiments. The sites Fe$_{12}$-Fe$_{14}$ are obtained from Fe$_{11}$ ($0.1184(6)$, $-0.001(1)$, $-0.0243(7)$) and Fe$_{22}$-Fe$_{24}$ from Fe$_{21}$ ($0.9386(6)$, $0.296(1)$, $0.8568(6)$) by symmetry elements $2_{1y}$, $i$, and $2_{1y}i$. Angles $\theta$ and $\phi$, which describe the orientation of the iron magnetic moments, are defined with respect to the $a^*bc$ coordinate system. Additionally, each spin has individual phase $\psi_{kl}$ [deg], where index $k$ =1, 2 counts the sites and the second index $l$ = 1-4 counts the atoms within the site.}
\label{tab2}
\begin{ruledtabular}
\begin{tabular}{ccccccc}
&$\theta$&$\phi$&$\psi_{11}$&$\psi_{12}$&$\psi_{13}$&$\psi_{14}$\\
{ Fe$_{11-14}$} &100(1)&-52(3)&{ 0 }&{ 55(5)}&{ 17(4)}&{ 260(10)}\\
& & &$\psi_{21}$&$\psi_{22}$&$\psi_{23}$&$\psi_{24}$\\
{ Fe$_{21-24}$} &100(1)&-45(3)&{ 10(5)}&{ 113(5)}&{ -10(11)}&{ 274(10)}\\
\end{tabular}
\end{ruledtabular}
\end{table}
\begin{figure}[b]
\includegraphics[width=65mm,keepaspectratio=true]{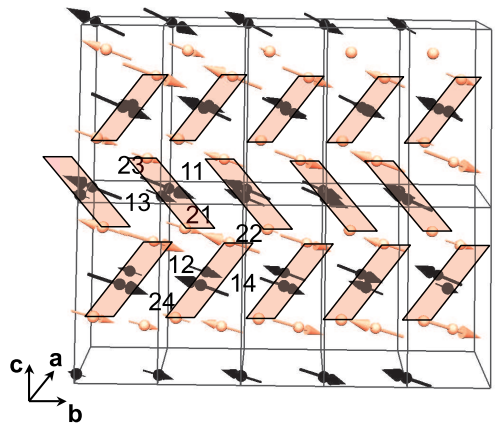}
\caption {Low temperature magnetic structure. Two different colors of arrows are used for the two sites, the Fe$^{3+}$ ions are labeled as in table~\ref{tab2}, the tetramers are shown schematically.}
\label{fig2}
\end{figure}
The combined refinement of polarization components and integrated magnetic intensities (25 and 41 independent reflections, respectively) using the CCSL code \cite{CCSL} yields excellent agreement between the experimental and calculated quantities (Fig.~\ref{fig1}). The best solution is the amplitude modulated model $S(i,k,l)=S_0\cos({\bf{q}}\cdot{\bf{r}}_{i} + \psi_{kl})$ with ${\bf{r}}_i$ being the vector defining the origin of the $i$-th cell. The modulation amplitude $S_0=4.02(9)\, \mu_B$ is the same for all iron sites in the unit cell, though each atom has its individual phase $\psi_{kl}$ (Table~\ref{tab2}). Magnetic moments on the same site in adjacent cells are collinear (Fig.~\ref{fig2}) and almost orthogonal to the wave vector $\bf{q}$, but their directions on Fe1 and Fe2 sites are inclined at a small  angle 7(3) deg (Table~\ref{tab2}).
There are two equally populated domains related by the $2_{1y}$ axis. We note that the incommensurate long-range magnetic order in \FeTe is most probably due to competing interactions within the geometrically frustrated iron tetramers.

Evidently, the magnetic structure has no inversion center. This removes the symmetry restriction for the coexistence of ferroelectric and magnetic order. We therefore decided to measure the temperature dependence of $\epsilon$ and spontaneous polarization $P$. An extremely sharp peak in $\epsilon'$ at $T_N=10.5(1)\, {\rm K}$ (Fig. 3a) announces a transition to a long-range ferroelectric state. At the same time, $\epsilon''$ is very small and frequency independent, proving intrinsic nature of the observed transition. The ferroelectric state is unambiguously confirmed by the emergence of $P$ at $T_N$ (Fig. 3b) and its reversal with the electric field (inset to Fig. 3a).
$P$ is the largest along the crystal $c-$axis, $P (c) = 8.5(2) \mu {\rm C/m}^2$. It is almost an order of magnitude smaller along $a^*$, $P (a^*) = 1.0(1)\,  \mu {\rm C/m}^2$, while for the $b$ direction it is below the sensitivity of our experimental equipment.
Comparing the temperature dependence of $P$ to the intensity of the magnetic ($\half$, 1.537, 0) peak, $I$, it is obvious that the two transitions coincide precisely (Fig. 3b).
{ When applying the magnetic field along the $a^*$ direction both the Neel-transition and the ferroelectric-transition temperatures simultaneously decrease to $T_N=9.4(3)$ K in the 9 T magnetic field.  This strong magnetic filed dependence  provide additional evidence for the magnetoelectric coupling in \FeTe.}

{ The phenomenological explanation for the occurrence of magnetoelectric effect in incommensurate helical or spiral magnetic phases has been given with thermodynamic potential terms type ${\bf{P}}\cdot\left[{\bf{M}}(\nabla\cdot\bf{M})-({\bf{M}}\cdot\nabla)\bf{M}\right]$ \cite{Most}. For our magnetic structure (Table II) we calculate that $P$ should lay in the $ab$ plane in striking contrast to the experimentally observed $P(c)$ component. We next extended calculations by additional ${\bf{P}}\cdot\nabla \left({\bf{M}}^2\right)$ term, which is important when $P$ is a sum of homogeneous and spatially modulated contributions \cite{Bet}. However, this additional term also cannot reproduce the correct ${\bf{P}}$ direction. Hence, it appears that  coupling terms, which work very well for helical or spiral magnetic orderings, cannot explain the appearance and the correct direction of the ferroelectric polarization in FeTe$_2$O$_5$Br. }

\begin{figure}[t]
\includegraphics[width=70mm,keepaspectratio=true]{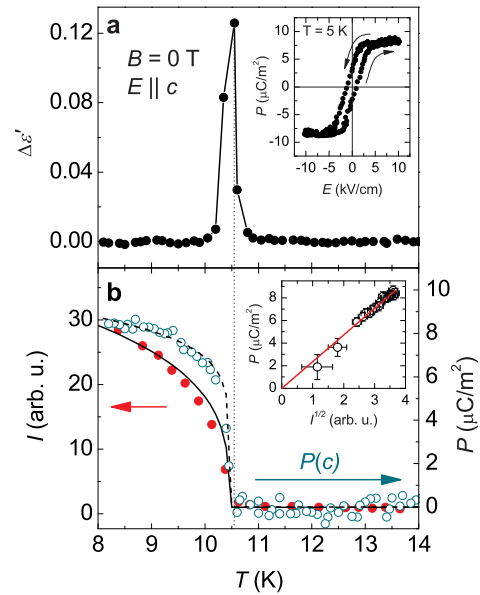}
\caption {(a) Temperature dependence of the change in the dielectric constant $\Delta\epsilon'=\epsilon' (T)-\epsilon' (14 K)$ measured for $E||c$. Inset: Ferroelectric hysteresis loop measured at $T=5$ K. (b) Temperature dependence of the spontaneous electric polarization, $P$, for $E||c$ (open circles, right scale) and the intensity of the ($\half$, 1.537, 0) neutron diffraction magnetic peak, $I$ (solid circles, left scale). { $I$ and $P$ calculated from Eq. (1) are presented with solid and dashed line respectively for $\beta$ = 0.15. Inset: A linear correlation between $\sqrt{I}$ and $P$. }
}
\label{fig3a}
\end{figure}

\begin{table}[t]
\caption{ Results of representation analysis for ${\bf{q}}$=($\half$, 0.463, 0) in $P 2_{1}/c$.
Top: Irreducible representations (IRR), bottom: complex basis vectors of magnetic moments for atoms 1 (x, y, z) and 2 (-x, y+1/2, -z+1/2) from the same orbit.
$\eta=cos(\pi q_y), \epsilon=sin(\pi q_y)$.
\label{tab3}}
\begin{ruledtabular}
\begin{tabular}{ccc}
IRR&(1$\mid$0)&($2_{1y}\mid00\half$)\\
$\Gamma_1$&1&$\eta$\\
$\Gamma_2$&1&-$\eta$\\
\end{tabular}
\begin{tabular}{cccccccc}
Irrep&Atom&\multicolumn{3}{c}{Re}&\multicolumn{3}{c}{Im}\\
$\Gamma_1$&1&1 0 0&{0 1 0}&0 0 1&0 0 0&0 0 0&0 0 0\\
&2&-$\eta$ 0 0& {0 $\eta$ 0}&0 0 -$\eta$&$\epsilon$ 0 0&0 -$\epsilon$ 0&0 0 $\epsilon$\\
$\Gamma_2$&1&{1 0 0}&0 1 0&0 0 1&0 0 0&0 0 0&0 0 0\\
&2&{$\eta$ 0 0}&0 -$\eta$ 0&0 0 $\eta$&-$\epsilon$ 0 0&0 $\epsilon$ 0&0 0 -$\epsilon$\\
\end{tabular}
\end{ruledtabular}
\end{table}

In order to better understand the magnetoelectric coupling in \FeTe we have performed representation analysis. The star of the wave vector is formed by the two vectors ${\bf{q}}$ and $-{\bf{q}}$, defining the little group, which is composed of two elements: identity $1$ and two-fold screw axis $2_{1y}$. It has two one-dimensional irreducible representations, $\Gamma_1$ and $\Gamma_2$ and the 4($e$) sites split into two orbits (Table~\ref{tab3}). Since the refined phase shift between the two magnetic moments from the same orbit (Table~\ref{tab2}) differs from the $\pi q_y$ = 83 deg value expected from the symmetry relations,  we conclude that our magnetic model is a combination of both $\Gamma_1$ and $\Gamma_2$.
{ The important coupling term, which already takes into account observed orientations of ${\bf{P}}$ and Fe$^{3+}$ magnetic moments as well as the symmetry operations of the little group, is written as
\begin{equation}
V=i\sum_{\alpha ,\beta}\varepsilon_{\alpha \beta}\left( S_\alpha ({\bf{q}},1)S^*_\beta ({\bf{q}},2)-
S^*_\alpha ({\bf{q}},1)S_\beta ({\bf{q}},2)\right) P_c\, .
\end{equation}
Here $\varepsilon_{\alpha \beta}$ is the magnetoelectric coupling tensor, $\alpha, \beta =x,y$ and $S_\alpha ({\bf{q}},i)$ is the Fourier component of the magnetic moments for Fe atoms $i=1,2$ (Table III).
For each irreducible representation we define a complex magnetic order parameter, whose magnitude in the vicinity of the phase transition can be described with the simple power law ansatz $(T_N-T)^\beta$. Phase difference between the two order parameters define the phases of individual amplitude modulation waves  $\psi_{kl}$ (Table II). The temperature dependence of $I$ and $P$  is simulated (Fig. 3b) by assuming temperature dependent $\psi_{kl}$ approaching low-temperature values obtained from the neutron diffraction experiments. The agreement with the experiment is much worse, if $\psi_{kl}$ are kept constant. The above analysis suggests that  sliding of the individual amplitude modulation waves, which also removes the center of inversion at the magnetic phase transition, is responsible for the magnetoelectric effect in FeTe$_2$O$_5$Br.}
Opposed to the $P\propto I$ dependence reported for representative magnetically incommensurate systems \cite{Fox, Yasui, Kenzelmann07} we find here the unusual proportionality between $\sqrt{I}$ and $P$ (inset to Fig. 3b). { Similar dependence in the low-temperature incommensurate spiral phase of Ni$_{3}$V$_{2}$O$_{8}$ \cite{NiVO} was explained with the saturation of the high-temperature magnetic order parameter already in the paraelectric phase.
In contrast, the observed $P\propto \sqrt{I}$  scaling in \FeTe is reproduced within our model as a direct consequence of the temperature dependence of the amplitude modulation wave phases.}

\begin{figure}[tb]
\includegraphics[width=76mm,keepaspectratio=true]{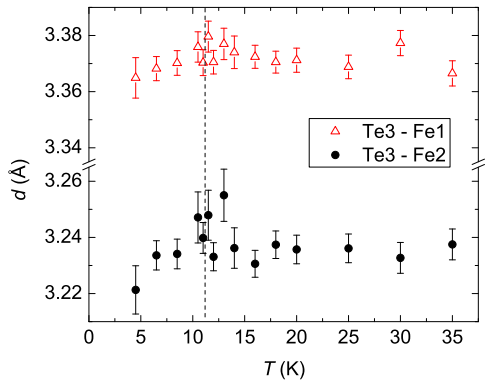}
\caption {Variation of the selected interatomic distances in the temperature range 4.5 K - 35 K from single crystal x-ray  diffraction. The labeling of the atoms corresponds to Ref.\onlinecite{Becker}.}
\label{fig4}
\end{figure}

To shed some additional light on microscopic picture of ferroelectricity and the magnetoelectric coupling we performed low-temperature single-crystal synchrotron X-ray diffraction experiments.
{ On cooling through the magnetic transition the deviations from the high-temperature crystallographic symmetry are very small and bellow the resolution of our XRD experiment.}
However, clearly distinguishable changes of the  Fe-Te interatomic distances (Fig.~\ref{fig4}) can be seen. This finding is important, because (i) Te$^{4+}$ ions bridge the intercluster exchange interactions and (ii) Te$^{4+}$ ions have lone-pair electrons. The observed structural anomalies therefore suggest the polarization of the Te$^{4+}$ lone-pair electrons and may thus explain the ferroelectricity in the magnetic phase. { We note that tetramer Fe-O interatomic distances also change slightly at the magnetic transition }
implying that the coupling between polar and magnetic order parameters is likely mediated through Fe-O-Te-O-Fe intercluster exchange.
{The standard spin-current \cite{SC} and "inverse Dzyaloshinskii-Moriya"  \cite{IDM}  models developed for spiral magnetic structures are unlikely to be active in FeTe$_2$O$_5$Br, since magnetic moments vary in amplitude and not in direction along ${\bf q}$.} Alternatively, exchange-striction model was frequently applied to magnetoelectrics with collinear  magnetic order \cite{ES,ES1,ES2,ES3}.
If exchange-striction model applies to \FeTe then the above  coupling term (Eq. (1)) suggests that the spin phonon coupling is provoked by the difference in the individual phases of  spin modulation waves. Additional experimental and theoretical investigations are necessary to validate this suggestion.

In summary, we have discovered simultaneous emergence of ferroelectric and magnetic order in FeTe$_2$O$_5$Br in the state with nearly transverse amplitude modulated incommensurate magnetic structure described by the wave vector $\bf{q}$=($\half$, 0.463, 0). The ferroelectricity is ascribed to the polarization of Te$^{4+}$ lone-pair electrons. { The magnetoelectric effect and the unusual temperature dependence of the magnetic and ferroelectric properties are explained with the sliding of neighbouring amplitude modulation waves opening the possibility for the exchange-striction in the Fe-O-Te-O-Fe intercluster exchange bridges.}   Our results suggest to look for new magnetoelectrics in the vast family of $M$-$T$-O-$X$ compounds ($M$ = Cu, Ni, Fe; $X$ = Cl, Br, I, $T$ = Te,  Se, Sb, Bi, Pb), as they frequently posses strong magnetic frustration  complemented by the presence of $T$ ions with lone-pair electrons.


We acknowledge fruitful discussions with  J.F. Scott and M. Kenzelmann. We thank Ya. Filinchuk and D. Chernyshov for settling up the x-ray diffraction experiment.  The sample preparation was supported by the NCCR research pool MaNEP of the Swiss NSF.


\begin{thebibliography}{99}
\bibitem{Kimura} T. Kimura et al., {\em Nature} {\bf 426}, 558 (2003).
\bibitem{Lottermoser} T. Lottermoser et al., {\em Nature} {\bf 430}, 541 (2004).
\bibitem{Nmat07} S. W. Cheong and M. Mostovoy, {\em Nature Mater.} {\bf 6}, 13 (2007).
\bibitem{Fiebig05} M. Fiebig, {\em J. Phys. D: Appl. Phys.} {\bf 38}, R123 (2005).
\bibitem{ScottNature} W. Eerenstein et al., {\em Nature} {\bf 442}, 759 (2006).
\bibitem{Kasuga} H. Katsura et al., {\em Phys. Rev. Lett.} {\bf 101}, 187207 (2008).
\bibitem{LonePair} R. Seshardi and N. A. Hill, {\em Chem. Mater.} {\bf 13}, 2892 (2001).
\bibitem{Becker} R. Becker et al., {\em J. Am. Chem. Soc.} {\bf 128}, 15469 (2006).
\bibitem{SHELXL} G. M. Sheldrick, { SHELXL97,} University of G\"{o}ttingen: G\"{o}ttingen, Germany, 1997.
\bibitem{eps1} Z. Kutnjak et al., {\em Nature} {\bf 441}, 956 (2006).
\bibitem{eps2} Z. Kutnjak and R. Blinc, {Phys. Rev.} {\bf B 76}, 104102 (2007).
\bibitem{CCSL} P. J. Brown, J. C. Matthewman, { CCSL}, 1897, (2008).
\bibitem{Fox} D. L. Fox et al. {\em Phys. Rev.} {\bf B 21}, 2926 (1980).
\bibitem{Yasui} Y. Yasui et al., {\em J. Phys. Soc. Jpn.} {\bf 77}, 023712 (2008).
\bibitem{Kenzelmann07} M. Kenzelmann et al., {\em Phys. Rev. Lett.} {\bf 98}, 267205 (2007).
\bibitem{Most} M. Mostovoy {\em Phys. Rev. Lett.} {\bf 96}, 067601, (2006).
\bibitem{Bet} J. J. Betouras et al. {\em Phys. Rev. Lett.} {\bf 98}, 257602, (2007).
\bibitem{NiVO} G. Lawes et al., {\em Phys. Rev. Lett.} {\bf 95}, 087205 (2005).
\bibitem{SC} H. Katsura et al., {\em Phys. Rev. Lett.} {\bf 95}, 057205 (2005).
\bibitem{IDM} I.A. Sergienko and E. Dagotto, {\em Phys. Rev.} {\bf B 73}, 094434 (2006).
\bibitem{ES} A. B. Harriset al., {\em Phys. Rev.} {\bf B 73}, 184433 (2006).
\bibitem{ES1} L. C. Chapon et al., {\em Phys. Rev. Lett.}, {\bf 93}, 177402 (2004).
\bibitem{ES2} N. Aliouane et al., {\em Phys. Rev.} {\bf B 73}, 020102(R) (2006).
\bibitem{ES3} I. A. Sergienko et al., {\em Phys. Rev. Lett.}, {\bf 97}, 227204 (2006).


\end{thebibliography}
\end{document}